\documentclass{ws-ijmpa}
\usepackage[super,compress]{cite}
\usepackage{url}
\usepackage{besphysics}
\usepackage{ulem}
\usepackage{color}
\usepackage{cancel}
\begin{document}

\markboth{Yinghua Tan, Zhenyu Zhang, Xiang Zhou}{Electromagnetic Dalitz Decays of $D_{(s)}^\ast$ Mesons}

\catchline{}{}{}{}{}

\title{Electromagnetic Dalitz Decays of $D_{(s)}^\ast$ Mesons}

\author{Yinghua Tan}

\address{Hubei Nuclear Solid Physics Key Laboratory, School of Physics and Technology, \\
Wuhan University, Wuhan, Hubei 430072, P. R. China\\
Department of Engineering Physics, Tsinghua University, \\
30 Shuangqing Road, Beijing 100084, P. R. China\\
tanyh@whu.edu.cn\\
tanyh21@mails.tsinghua.edu.cn}

\author{Zhenyu Zhang$^\ast$\textsuperscript{, }\footnotemark[3]$\ $ and Xiang Zhou$^\dagger$\textsuperscript{, }\footnotemark[3]}

\renewcommand{\thefootnote}{\fnsymbol{footnote}}
\footnotetext[3]{Correspoding authors.}
\renewcommand{\thefootnote}{\alph{footnote}}

\address{Hubei Nuclear Solid Physics Key Laboratory, School of Physics and Technology, \\
Wuhan University, Wuhan, Hubei 430072, P. R. China\\
$^\ast$zhenyuzhang@whu.edu.cn\\
$^\dagger$xiangzhou@whu.edu.cn}



\maketitle

\begin{history}
\received{Day Month Year}
\revised{Day Month Year}
\end{history}

\begin{abstract}
Rare electromagnetic decays of charmed mesons are useful laboratories to explore the structure of hadronic states and the interactions between photons and charmed mesons, to test the chiral perturbation theory in flavor sector and to search for new physics including dark photons. In this paper, we calculate the relative branching ratios of electromagnetic Dalitz decays $D_{(s)}^\ast\to D_{(s)}\LL$ to their corresponding radiative decays $D_{(s)}^\ast\to D_{(s)}\gamma$ with transition form factor in vector-meson dominance model, where $D_{(s)}^\ast$ represents $D^\ast(2007)^0$, $D^\ast(2010)^\pm$, $D^\ast(2640)^\pm$, $D_s^{\ast\pm}$, $D_{s1}^\ast(2700)^\pm$ and $D_{s1}^\ast(2860)^\pm$. The corresponding dileptonic invariant mass spectra and the leptonic angular distributions are also presented. 
\end{abstract}


\ccode{PACS numbers:12.39.Fe, 12.40.Vv, 13.25.Ft, 13.40.Gp.}


\section{Introduction}
The electromagnetic (EM) decays of mesons have attracted much attention for their rich physical phenomenons\cite{odonnell1981}. The EM radiative decays of light vector mesons, such as $\rho^0\to\piz\gam$ and $\phi\to\eta\gam$, provide opportunities to study the couplings between photons and mesons. The investigations of the couplings give a deeper insight into the mixture of $\omega-\phi$ and $\eta-\etap$, SU(3) flavor symmetry breaking\cite{odonnell1981, bramon2001}. With the discovery of the fourth quark, charm quark, much more EM radiative decays were discovered, such as $\jpsi\to\eta_c(1S)\gam$\cite{baltrusaitis1984}, $\jpsi\to\etap\gam$\cite{bartel1976} and so on. The couplings between photons and mesons can be studied in much richer decays. If a vector meson emits an off-shell photon which turns into dilepton, such as $\rho^0\to\piz\EE$, $\phi\to\eta\MM$, EM Dalitz decays arise. EM Dalitz decays can be used to study the structures of the mesons. After the dark photon, which explains dark matter\cite{febbrichesi2020}, was carried out, it was suggested to search for dark photon in EM decays, such as $\piz/\eta/\etap\to\gam A^\prime$\cite{meijer1992, meijer1994, amsler1994, altegoer1998, gninenko2012, adlarson2013, agakishiev2014, adare2015, gninenko2012a}, where $A^\prime$ represents the dark photon. EM Dalitz decays of the charmed sector, such as $D^{\ast0}\to D^0\EE$ and $\jpsi\to P\LL$, are also suggested to search for it\cite{fu2012, ilten2015}, where $P$ represents pseudoscalar mesons and $\ell$ leptons. $D^{\ast+}\to D^+\EE$ and $D_s^{\ast+}\to D_s^+\EE$ are proposed as sensitive channels to search for light vector meson $X(17)$\cite{castro2021}, which is relative to anomalies in $^8\mathrm{Be}^\ast$ and $^4\mathrm{He}$ dileptonic transition.

The relative branching ratios of EM Dalitz decays for unflavored vector mesons to those of their corresponding EM radiative decays have been measured\cite{zyla2020} and can be theoretically described by transition form factors(TFFs). The TFF can be expressed as $\mathcal{F}(q^2)=1/(1-q^2/\Lambda^2)$\cite{landsberg1975a, landsberg1975b}in the simple pole approximation, where $q$ is the four-momentum of dilepton and $\Lambda$ is the pole mass. The pole mass can be determined in experiments by using $\Lambda^{-2}=(\mathrm{d}\mathcal{F}(q^2)/\mathrm{d}q^2)_{q^2=0}$. For example, N60  and SND experiments measured the pole masses of $\omega\to\piz\EE$\cite{arnaldi2009} and $\phi\to\eta\MM$\cite{achasov2001}, respectively. Vector-Meson Dominance (VMD) model is used to describe the interaction between photons and mesons and it gives $\Lambda\approx m_{V^\prime}$, where $V^\prime$ is the dominant vector meson mediator. This approximation is useful when predicting the EM Dalitz decays of vector mesons\cite{landsberg1975a, landsberg1975b}, including charmonia\cite{fu2012, gu2019, zhang2019} and bottomia\cite{gu2019}.

$D_s^{\ast\pm}\to D_s^\pm\EE$ is also an EM Dalitz decay belonging to the charmed strange vector meson sector. Its branching ratio was measured to be $(7.2^{+1.5}_{-1.3}\pm1.0)\times10^{-3}$\cite{cronin-hennessy2012}. The relative branching ratio of it to that of EM radiative decay was predicted to be $0.64\%$ by using the TFF $1/(1-q^2/m_\phi^2)$ and it was consistent with the experimental result\cite{cronin-hennessy2012}. The relative branching ratio of $D^{\ast0}\to D^0\EE$ was predicted to be $0.64\%$\cite{ilten2015}. Besides $D_s^{\ast\pm}$ and $D^{\ast0}$, charmed vector mesons also includes $D^{\ast\pm}$ and their excited state particles, $D^\ast(2640)^\pm$, $D_{s1}^\ast(2700)^\pm$ and $D_{s1}^\ast(2860)^\pm$. 

In this paper, we calculate the relative branching ratios of $D_{(s)}^\ast\to D_{(s)}\LL$ to $D_{(s)}^\ast\to D_{(s)}\gam$ with TFFs in VMD model, and present the dileptonic invariant mass spectra and the angular distributions of lepton, where $D_{(s)}^\ast$ represents $D^\ast(2007)^0$, $D^\ast(2010)^\pm$, $D^\ast(2640)^\pm$, $D_s^{\ast\pm}$, $D_{s1}^\ast(2700)^\pm$ and $D_{s1}^\ast(2860)^\pm$
, $D_{(s)}$ denotes the charmed pseudoscalar meson $D^0$, $D^\pm$ and $D_s$, and $\ell$ denotes electron and muon. Sec. \ref{theory} gives the derivations of the relative differential decay rates. Sec. \ref{calculation} gives the relative branching ratios, invariant mass spectra and angular distributions. Sec. \ref{end} gives the summary and discussions.

\section{The relative differential decay rates of $D_{(s)}^\ast\to D_{(s)}\LL$}\label{theory}

The $\LL$ pair in the EM Dalitz decays $D_{(s)}^\ast\to D_{(s)}\LL$ is produced from $\gam$ conversion of the EM radiative decays $D_{(s)}^\ast\to D_{(s)}\gam^\ast$. The effective Lagrangian of the  interaction vertex $D_{(s)}^\ast-D_{(s)}-\gam$ is\cite{faessler2000}
\beq\label{Lagrangian}
\mathcal{L}_\mathrm{eff}=g_{D_{(s)}^\ast\gamma D_{(s)}}\epsilon^{\alpha\beta\gam\delta}(\partial_\alpha V_{D_{(s)}^\ast\beta} \partial_\gam A_{\delta})P_{D_{(s)}},
\eeq
where $g_{D_{(s)}^\ast VD_{(s)}}$ is the coupling between $D_{(s)}^\ast$, $D_{(s)}$ and vector particles, $V_{D_{(s)}^\ast}$, $A$ and $P_{D_{(s)}}$ are the $D_{(s)}^\ast$, $\gamma$ and $D_{(s)}$ fields, respectively. 

According to the VMD model, the decays $D_{(s)}^\ast\to D_{(s)}\LL$ are proceed dominantly by $D_{(s)}^\ast\to D_{(s)}V^\ast\to D_{(s)}\gam^\ast\to D_{(s)}\LL$, where $V^\ast$ is a virtual vector meson $\rho^0$, $\omega$ or $\phi$. The $q^2$-dependent TFFs $f(q^2)$ can be written as\cite{cheng2021}
\beqr
\label{DstarpmDpmgam}
f_{D^{\ast\pm}}(q^2)&=&\frac{1}{\sqrt{2}}[g_{D^{\ast\pm}\omega D^\pm}\frac{e}{2g_{\omega\gam}}\mathcal{G}_\omega(q^2)+g_{D^{\ast\pm}\rho^0D^\pm}\frac{e}{2g_{\rho\gam}}\mathcal{G}_\rho(q^2)]\\
\label{Dstar0D0gam}
   f_{D^{\ast0}}(q^2)&=&\frac{1}{\sqrt{2}}[g_{D^{\ast0}\omega D^0}\frac{e}{2g_{\omega\gam}}\mathcal{G}_\omega(q^2)+g_{D^{\ast0}\rho^0D^0}\frac{e }{2g_{\rho\gam}}\mathcal{G}_\rho(q^2)]\\
\label{DsstarpmDspmgam}
   f_{D_s^{\ast\pm}}(q^2)&=&g_{D_s^{\ast\pm}\phi D_s^\pm}\frac{e}{2g_{\phi}}R_\phi\mathcal{G}_\phi(q^2)
\eeqr
where $g_{V\gam}$ is the coupling between vector meson and photon, $e=\sqrt{4\pi\alpha}$, $R_\phi=m_{u/d}/m_s$, and $\mathcal{G}_V(q^2)=1/(1-q^2/m_V^2-i\Gamma_V(q^2)/m_V)$.

The couplings between vector mesons and photons can be determined by $V\to\EE$ decays. The decay widths of $V\to\EE$ is\cite{faessler2000}
\beq
\Gamma(V\to\EE)=\frac{2\pi\alpha^2}{3g_{V\gam}^2}(1+\frac{2m_e^2}{m_V^2})|\boldsymbol{p}_e|
\eeq
where $m_e$ and $m_V$ are the masses of electron and vector mesons, $|\boldsymbol{p}_e|$ is the magnitude of momentum of outgoing electron. According to the quark model, the inverse of couplings $g_{V\gamma}$ are proportional to the charge of component quarks in vector mesons, and it gives $g_{\rho^0\gamma}:g_{\omega\gamma}:g_{\phi\gamma}=1:3:-3/\sqrt{2}$ in SU(3) symmetry\cite{faessler2000}. The $g_{V\gam}$'s of $\rho^0$, $\omega$ and $\phi$ are listed on the Table \ref{gVgamma}.
\btbl[!htb]
\tbl{The value of coupling constants between photons and vector mesons $g_{V\gamma}(V=\rho^0, \omega, \phi)$ extracted from data\cite{zyla2020}. \label{gVgamma}}
{\begin{tabular}{@{}cccc@{}} \toprule
decay channels &	$\rho^0\to\EE$	&	$\omega\to\EE$	&	$\phi\to\EE$	\\\colrule
$\Gamma(V\to\EE)(\mathrm{KeV})$	&	$6.98\pm0.09$	&	$0.625\pm0.014$	&	$1.22\pm0.01$	\\
$|\boldsymbol{p}_e|(\mathrm{MeV})$	&	$387.63\pm24.23$	&	$391.32\pm5.64$	&	$509.73\pm0.13$	\\
$g_{V\gam}$	&	$2.49\pm0.08$	&	$8.36\pm0.11$	&	$-6.83\pm0.04$	\\\botrule
\end{tabular}}
\etbl

The couplings between $D_{(s)}^\ast$, $D_{(s)}$ and vector mesons can be expressed by a coupling constant $g_{VVP}$. The relations between them are\cite{tornqvist1979}:
\beqr
g_{VVP}&=&-\sqrt{2}g_{D^{\ast\pm}\rho^0D^\pm}=\sqrt{2}g_{D^{\ast\pm}\omega D^\pm}\nonumber\\
g_{VVP}&=&\sqrt{2}g_{D^{\ast0}\rho^0D^0}=\sqrt{2}g_{D^{\ast0}\omega D^0}\nonumber\\
g_{VVP}&=&g_{D_s^{\ast\pm}\phi D_s^\pm}\nonumber
\eeqr

The couplings of other vector-vector-pseudoscalar$(V-V-P)$ vertex can also be written by $g_{VVP}$. The EM radiative decays of vector mesons or pseudoscalar mesons that are measured can also be presented by $g_{VVP}$ and $g_{V\gam}$. By fitting the EM radiative decays, $g_{VVP}$ is obtained as $-8.15\pm0.11\gev^{-1}$, which is consistent with that in Ref. \refcite{cheng2021}. The coupling equations and fitted results are shown on the Table \ref{gVVPfitting}. $R_\phi=0.8$ is used in this fitting process.
\btbl[!htb]
\tbl{The $q^2$-dependent TFF of $V\to P\gam$ or $P\to V\gam$ decays and the fitted decay widths.\label{gVVPfitting}}
{\begin{tabular}{cccc} \toprule
decay channels					&	$\Gamma(\mathrm{KeV})$\cite{zyla2020}	&	$q^2$-dependent TFF	&	$\Gamma_\mathrm{fit}(\mathrm{KeV})$	\\	\colrule
$\rho^0\to\piz\gamma$			&	$69.47\pm8.88$			&	$g_{\rho^0\omega\piz}\frac{e}{2g_{\omega\gam}}\mathcal{G}_\omega(q^2)$			&	$61.36\pm1.54$		\\
$\rho\to\eta\gamma$				&	$44.34\pm3.12$			&	$g_{\rho^0\rho^0\eta}\frac{e}{2g_{\rho\gam}}\mathcal{G}_\rho(q^2)$				&	$41.40\pm1.04$		\\
$\etap\to\rho\gamma$			&	$55.46\pm1.92$			&	$g_{\rho^0\rho^0\etap}\frac{e}{2g_{\rho\gam}}\mathcal{G}_\rho(q^2)$				&	$99.42\pm2.50$		\\
$\omega\to\piz\gamma$			&	$713.16\pm19.85$		&	$g_{\omega\rho^0\piz}\frac{e}{2g_{\rho\gam}}\mathcal{G}_\rho(q^2)$				&	$713.01\pm17.91$	\\
$\omega\to\eta\gamma$			&	$3.82\pm0.34$			&	$g_{\omega\omega\eta}\frac{e}{2g_{\omega\gam}}\mathcal{G}_\omega(q^2)$			&	$4.00\pm0.10$		\\
$\etap\to\omega\gamma$			&	$4.74\pm0.20$			&	$g_{\omega\omega\etap}\frac{e}{2g_{\omega\gam}}\mathcal{G}_\omega(q^2)$			&	$7.89\pm0.20$		\\
$\phi\to\eta\gamma$				&	$55.36\pm1.08$			&	$g_{\phi\phi\eta}\frac{e}{2g_{\phi\gam}}R_\phi\mathcal{G}_\phi(q^2)$			&	$59.65\pm1.50$		\\
$\phi\to\etap\gamma$			&	$0.26\pm0.01$			&	$g_{\phi\phi\etap}\frac{e}{2g_{\phi\gam}}R_\phi\mathcal{G}_\phi(q^2)$			&	$0.21\pm0.01$		\\
$\rho^\pm\to\pipm\gamma$		&	$67.10\pm7.46$			&	$g_{\rho^\pm\omega\pipm}\frac{e}{2g_{\omega\gam}}\mathcal{G}_\omega(q^2)$		&	$60.93\pm1.53$		\\
$\kaon^{*\pm}\to\kaonpm\gamma$	&	$50.29\pm4.66$			&	\footnotemark[1]	&	$24.57\pm0.62$		\\
$\kaon^{*0}\to\kaon^0\gamma$	&	$116.36\pm10.01$		&	\footnotemark[2]	&	$79.79\pm2.00$		\\
$D^{*\pm}\to D^\pm\gamma$		&	$1.33\pm0.33$			&	Eq. \eqref{DstarpmDpmgam}		&	$2.01\pm0.05$		\\
$D^{*0}\to D^0\gamma$			&	$<741.30$				&	Eq. \eqref{Dstar0D0gam}			&	$7.05\pm0.18$		\\
$D_s^{\ast\pm}\to D_s^\pm\gam$	&	$<1776.50$				&	Eq. \eqref{DsstarpmDspmgam}		&	$1.49\pm0.04$	\\ \botrule
\end{tabular}}
\footnotesize{$Notes$: \textsuperscript{a}$\frac{1}{\sqrt{2}}[g_{\kaon^{\ast\pm}\omega\kaon^\pm}\frac{e}{2g_{\omega\gam}}\mathcal{G}_\omega(q^2)+g_{\kaon^{\ast\pm}\rho^0\kaon^\pm}\frac{e}{2g_{\rho\gam}}\mathcal{G}_\rho(q^2)]+g_{D^{\ast\pm}\phi D^\pm}\frac{e}{2g_{\phi\gam}}R_{\phi}\mathcal{G}_{\phi}(q^2)$.\\
\textsuperscript{b}$\frac{1}{\sqrt{2}}[g_{\kaon^{\ast0}\omega\kaon^0}\frac{e}{2g_{\omega\gam}}\mathcal{G}_\omega(q^2)+g_{\kaon^{\ast0}\rho^0\kaon^0}\frac{e}{2g_{\rho\gam}}\mathcal{G}_\rho(q^2)]+g_{D^{\ast\pm}\phi D^\pm}\frac{e}{2g_{\phi\gam}}R_{\phi}\mathcal{G}_{\phi}(q^2)$.}
\etbl

The  widths of $\omega$ and $\phi$ are narrow, so $\Gamma_\omega(q^2)$ and $\Gamma_\phi(q^2)$ are adopted as the nominal widths of $\omega$ and $\phi$, i. e. $\Gamma_{\omega/\phi}(q^2)=\Gamma_{\omega/\phi}$. But the width of $\rho^0$ is very wide, so $\Gamma_{\rho^0}(q^2)$ has the form\cite{qin2018}:
\beq
\Gamma_{\rho^0}(q^2)=\Gamma_{\rho^0}\frac{\sqrt{q^2}}{m_{\rho^0}}(\frac{1-4m_\pi^2/q^2}{1-4m_\pi^2/m_{\rho^0}^2})^{3/2}\Theta(q^2-4m_\pi^2),
\eeq
where $\Gamma_{\rho^0}$ is the nominal width of $\rho^0$, and $m_\pi$ is the mass of $\pi^\pm$.

With the coupling constants $g_{D_{(s)}^\ast VD_{(s)}}$, $g_{V\gamma}$ and width $\Gamma_V(q^2)$ determined, the widths of radiative decays and EM Dalitz decays of $D_{(s)}^\ast$ can be obtained. In the decays $D_{(s)}^\ast\to D_{(s)}\gam$, the decay widths of $D_{(s)}^\ast\to D_{(s)}\gam$ can be derived from Eq. \eqref{Lagrangian}-\eqref{DsstarpmDspmgam} as\cite{fu2012}
\beq\label{VPgamma}
\Gamma_{D_{(s)}\gam}=\frac{1}{3}\frac{f^2(0)}{4\pi}(\frac{m_{D_{(s)}^\ast}^2-m_{D_{(s)}}^2}{2m_{D_{(s)}^\ast}})^3,
\eeq
where $m_{D_{(s)}^\ast}$ and $m_{D_{(s)}}$ represent the masses of $D_{(s)}^\ast$ and $D_{(s)}$, respectively.  $q^2=0$ here since the photon is on shell.

The differential decay widths of EM Dalitz decays $D_{(s)}^\ast\to D_{(s)}\LL$ can also be obtained as\cite{fu2012}
\beqr\label{VPEE}
\mathrm{d}\Gamma_{D_{(s)}\LL}&=&\frac{1}{3}\frac{\alpha}{2^4(2\pi)^4}\frac{f^2(q^2)}{q^2}\sqrt{1-\frac{4m_{\ell}^2}{q^2}}(\frac{\sqrt{\lambda(m_{D_{(s)}^\ast}^2, m_{D_{(s)}}^2, q^2)}}{2m_{D_{(s)}^\ast}})^3\nonumber\\
&&\times[(1+\frac{4m_{\ell}^2}{q^2})+(1-\frac{4m_{\ell}^2}{q^2})\cos^2\theta^\ast]\mathrm{d}q^2\mathrm{d}\Omega^\ast\mathrm{d}\Omega,
\eeqr
where the fine structure constant $\alpha\approx1/137$, $m_{\ell}$ is the mass of lepton, the K\"{a}llen function $\lambda(a, b, c) = a^2+b^2+c^2-2ab-2bc-2ca$, $\theta^\ast$ is the polar angle of positive lepton in the rest frame of virtual photon, $\mathrm{d}\Omega^\ast$ and $\mathrm{d}\Omega$ are the solid angle of positive lepton in the rest frame of virtual photon and the solid angle of $D_{(s)}$ meson in the rest frame of $D_{(s)}^\ast$ meson, respectively.

The differential relative branching ratio of $D_{(s)}^\ast\to D_{(s)}\LL$ to $D_{(s)}^\ast\to D_{(s)}\gam$ is obtained from Eq. \eqref{VPgamma} and Eq. \eqref{VPEE}\cite{landsberg1975a, landsberg1975b}, which is
\beq\label{RBR}
\frac{\mathrm{d}\Gamma_{D_{(s)}\LL}}{\Gamma_{D_{(s)}\gam}\mathrm{d}q^2} = \frac{\alpha}{3\pi}\frac{|F(q^2)|^2}{q^2}\sqrt{1-\frac{4m_{\ell}^2}{q^2}}(1+\frac{2m_{\ell}^2}{q^2})(\frac{\sqrt{\lambda(m_{D_{(s)}^\ast}^2, m_{D_{(s)}}^2, q^2)}}{m_{D_{(s)}^\ast}^2-m_{D_{(s)}}^2})^3,
\eeq
where $F(q^2)$ is the normalized TFF which is defined as\cite{landsberg1975a, landsberg1975b}:
\beq
F(q^2) = \frac{f(q^2)}{f(0)}.
\eeq

\section{Numerical Results}\label{calculation}

The curves of $F^2(q^2)$ can be a tool to study the TFF formula in experiments\cite{arnaldi2009, achasov2001, dzhelyadin1981}. Fig. \ref{FVMDfigure} shows the curves of $F^2(q^2)$ as functions of $q^2$ in the range of $[0, 0.9\gev]$, for $\Delta m = m_{D_{(s)}^\ast} - m_{D_{(s)}}$ is no more than $0.9\gev$, where $q^2=m_{\LL}^2$. All the square of the VMD model TFFs go to $1$ when $m_{\LL}$ goes to zero, and this property obeys the normalization requirement of normalized TFF. When $m_{\LL}$ gets bigger, it is obvious that $F^2_{D^\ast}(m_{\LL}^2)$ are different from $F^2_{D_s^\ast}(m_{\LL}^2)$ which is mainly caused by their pole mass positions. The pole positions of $F^2_{D^\ast}(m^2_{\LL})$ are at $m_\omega$ and $m_{\rho^0}$, while the pole position of $F^2_{D_s^\ast}(m^2_{\LL})$ is at $m_\phi$ which is outside of the permitted phase space range. If $\Delta m > m_V$, where $V$ is denoted as vector meson $\rho^0$, $\omega$ or $\phi$, the vector meson can be on shell and then a pole in the $\LL$ invariant mass spectra can appear. For example, for $D^\ast(2640)^\pm\to D^\pm\LL$, the $\Delta m$ is around 770 MeV, which is very close to $m_{\rho^0}$. Because of the broad width of  $\rho^0$ meson, it is expected that there are some enhancements in the $\LL$ invariant mass spectra around $m_{\rho^0}$ which can be used to check the corresponding VMD model TTF experimentally.

\bfg[!htb]
\centerline{\includegraphics[width=0.7\textwidth]{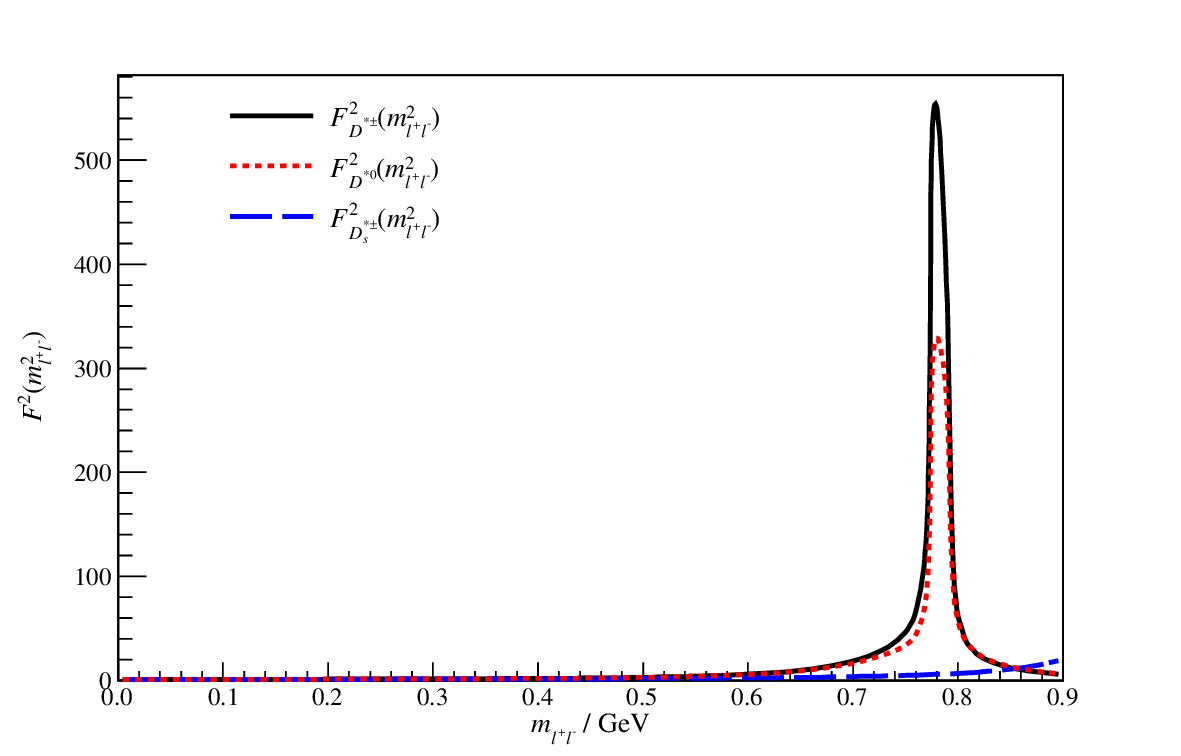}}
\caption{The curves of $F^2(m_{\LL}^2)$ versus $m_{\LL}$, where $m_{\LL}=\sqrt{q^2}$. The black solid line represents the square of TFF for $D^{\ast\pm}$, the red dotted line represents that for $D^{\ast0}$, and the blue dashed line represents that for $D_s^{\ast\pm}$.}\label{FVMDfigure}
\efg


The signs relative $\rho^0$ part and $\omega$ part are opposite in Eq. \eqref{DstarpmDpmgam} and Eq. \eqref{Dstar0D0gam} for $F_{D^{\ast\pm}}(q^2)$ and $F_{D^{\ast 0}}(q^2)$, respectively. Thus there are $\rho^0$, $\omega$ and the interference parts in $F^2_{D^{\ast\pm}}(q^2)$ and $F^2_{D^{\ast 0}}(q^2)$ which are shown in Fig. \ref{TFFDstarDivide}. The $\rho^0$ parts in both $F^2_{D^{\ast\pm}}(q^2)$ and $F^2_{D^{\ast 0}}(q^2)$ are broader but much less than the $\omega$ and  interference parts around $m_{\rho^0/\omega}$. The peak values of interference parts are about one-third for the ones of the $\omega$ parts with the negative sign for $F^2_{D^{\ast\pm}}(q^2)$ but positive sign for $F^2_{D^{\ast 0}}(q^2)$. The peak value of $\omega$ part in $F^2_{D^{\ast\pm}}(q^2)$ is about four times larger than the one in $F^2_{D^{\ast 0}}(q^2)$. Because of the opposite sign of the interference parts, the peak value of $F^2_{D^{\ast\pm}}(q^2)$ around $m_{\rho^0/\omega}$ is less than two times for the one of $F^2_{D^{\ast 0}}(q^2)$ so that form factors $F_{D^{\ast\pm}}(q^2)$ and $F_{D^{\ast 0}}(q^2)$ shown in Fig. \ref{FVMDfigure} are almost the same. Considering the narrow phase space of $D^{\ast}(2010)^\pm\to D^\pm\EE$ and $D^{\ast}(2007)^0\to D^0\EE$, their relative branching ratios are expected almost the same.

\bfg[!htb]
\begin{minipage}[c]{0.5\textwidth}
    \centerline{\includegraphics[width=\textwidth]{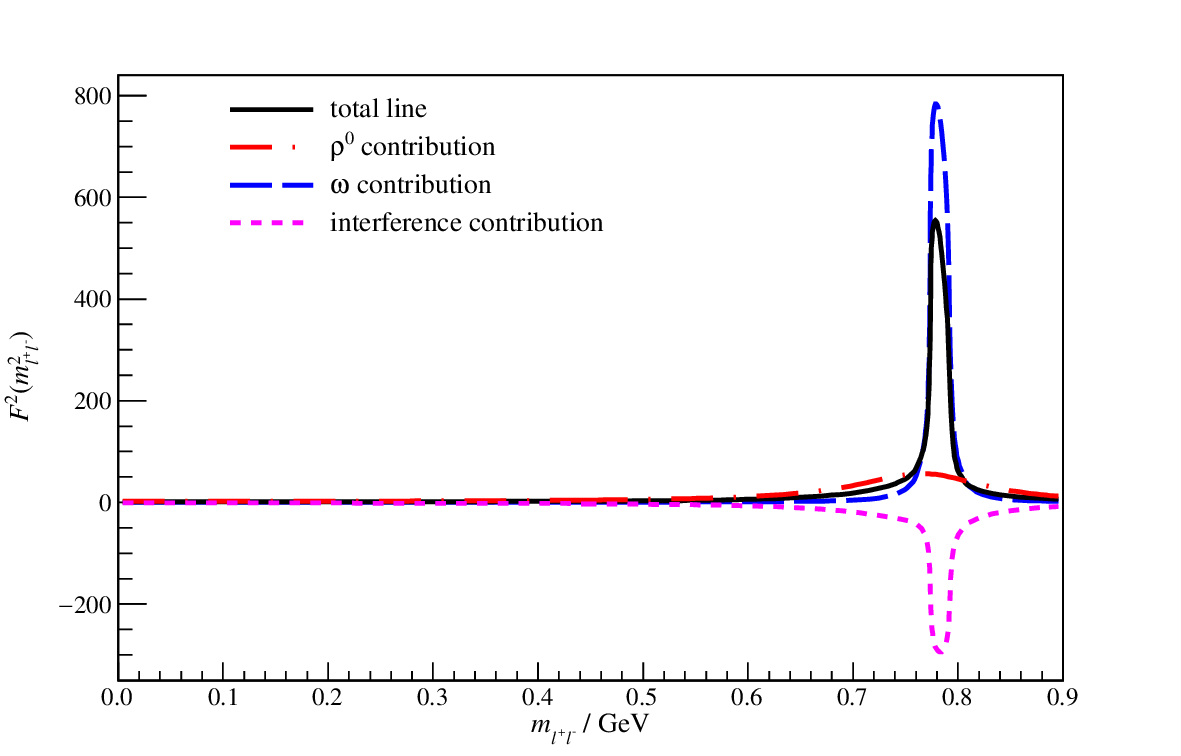}}
\end{minipage}%
\begin{minipage}[c]{0.5\textwidth}
    \centerline{\includegraphics[width=\textwidth]{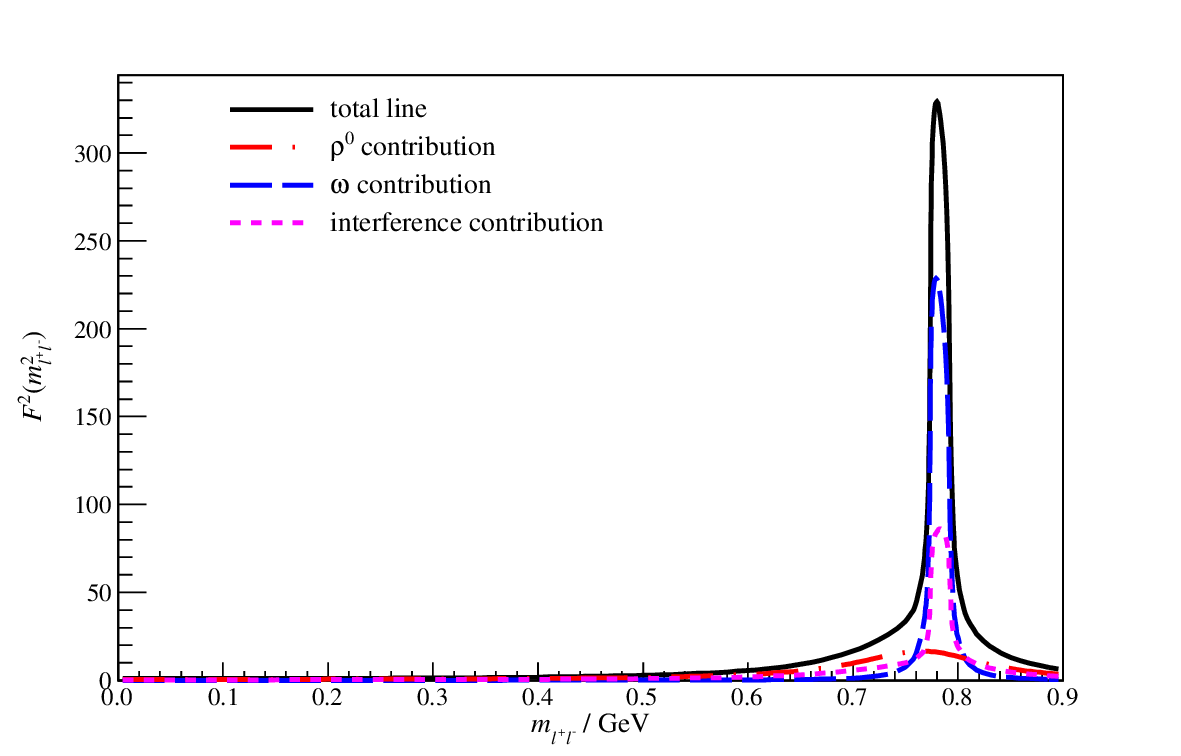}}
\end{minipage}
\caption{Contributions of $\rho^0$, $\omega$ and interference parts to TFFs for $D^{\ast\pm}$(left) and $D^{\ast0}$(right). The black solid line represents the square of total TFF, while the red dotted dashed line, the blue long dashed line and the pink short dashed line represent the contributions of $\rho^0$, $\omega$ and interference parts.}\label{TFFDstarDivide}
\efg

The relative branching ratios of $D_{(s)}^\ast\to D_{(s)}\LL$ to $D_{(s)}^\ast\to D_{(s)}\gam$ can be derived by integrating Eq. \eqref{RBR} over $q^2$ from $(2m_{\ell})^2$ to $(m_{D_{(s)}^\ast}-m_{D_{(s)}})^2$. The results are shown on Table \ref{RBRresults}. For the $\EE$ decays with small $\Delta m$, such as the EM Dalitz decays of $D^\ast(2007)^0$, $D^\ast(2010)^\pm$ and $D_s^\ast$, the relative branching ratios are about $0.6\%$, but for the decays with large $\Delta m$, they are from $0.9\%$ to $1.0\%$. This is because the phase spaces of the EM Dalitz decays of $D^\ast(2640)^\pm$, $D_{s1}^\ast(2700)^\pm$ and $D_{s1}^\ast(2860)^\pm$ are larger than those of $D^\ast(2010)^\pm$ and $D_s^\ast$ when they decay into final states with the same dilepton and TFFs. For the EM Dalitz decays with $\MM$ final states, the relative branching ratios are suppressed by the phase spaces. The relative branching ratios of the EM Dalitz decays between neutral $D^\ast(2007)^0$ and charged $D^\ast(2010)^\pm$ are almost the same. This shows that the decays obey the isospin symmetry. The predicted result of $D_s^\ast\to D_s\EE$ is the same as that in Ref. \refcite{cronin-hennessy2012}. The relative branching ratio of $D_s^{\ast+}\to D_s^+\EE$ to $D_s^{\ast+}\to D_s^+\gam$ on Particle Data Group(PDG) is $(0.72\pm0.17)\%$\cite{zyla2020}, which is consistent with the results on Table \ref{RBRresults}.

\btbl[!htb]
\tbl{The relative branching ratios of $D_{(s)}^\ast\to D_{(s)}\LL$ to $D_{(s)}^\ast\to D_{(s)}\gam$, derived by integrating the differential relative decay widths over the phase space.\label{RBRresults}}
{\begin{tabular}{@{}ccc@{}}   \toprule
Decay channels  &   $\ell=e(\%)$ &   $\ell=\mu(\%)$   \\  \colrule
$D^\ast(2007)^0\to D^0\LL$ 				& 0.645 &        \\ 
$D^\ast(2010)^\pm\to D^\pm\LL$ 			& 0.644 &        \\
$D_s^{\ast\pm}\to D_s^\pm\LL$\footnotemark[1] 			& 0.646 &        \\ 
$D^\ast(2640)^\pm\to D^\pm\LL$\footnotemark[1] 			& 1.034 & 0.205  \\
$D_{s1}^\ast(2700)^\pm\to D_s^\pm\LL$ 	& 0.943 & 0.122  \\ 
$D_{s1}^\ast(2860)^\pm\to D_s^\pm\LL$ 	& 1.003 & 0.178  \\ \botrule
\end{tabular}}
\begin{tabnote}
    \textsuperscript{a}$D_s^{\ast\pm}$ and $D^\ast(2640)^\pm$ are treated as vector mesons although their quantum numbers have not been determined yet\cite{zyla2020}.
\end{tabnote}
\etbl

The dileptonic invariant mass spectra are shown in Fig. \ref{massspectrum} for each decay channel. For the decay channels with small $\Delta m$, such as $D^\ast(2007)^0\to D^0\EE$, $D^\ast(2010)^\pm\to D^\pm\EE$ and $D_s^{\ast\pm}\to D_s^\pm\EE$, the phase spaces are small, so it is difficult to check the VMD model TFF from the electron-positron pair invariant mass spectra. But for the decays with large $\Delta m$, there are some enhancements as expected in the invariant mass spectra of dileption at the $\rho^0$ and $\omega$ positions for $D^\ast(2640)^\pm\to D^\pm\EE/\MM$ shown in Fig. \ref{massspectrum}, which can be used to check the VMD model TFF in experiments.  Since $m_e$ is much smaller than $m_\mu$, the electron-positron pair almost goes along the recoil direction of $D_{(s)}$ in the rest frame of $D_{(s)}^\ast$, so the invariant mass spectra shows a peak at the low energy region. The invariant mass spectra with dimuon final state are much different from those with electron-positron pair final state if one compares the spectra between $D_{(s)}^\ast\to D_{(s)}\EE$ and $D_{(s)}^\ast\to D_{(s)}\MM$.

\bfg[!htb]
\begin{minipage}{0.33\textwidth}
\centerline{\includegraphics[width=\textwidth]{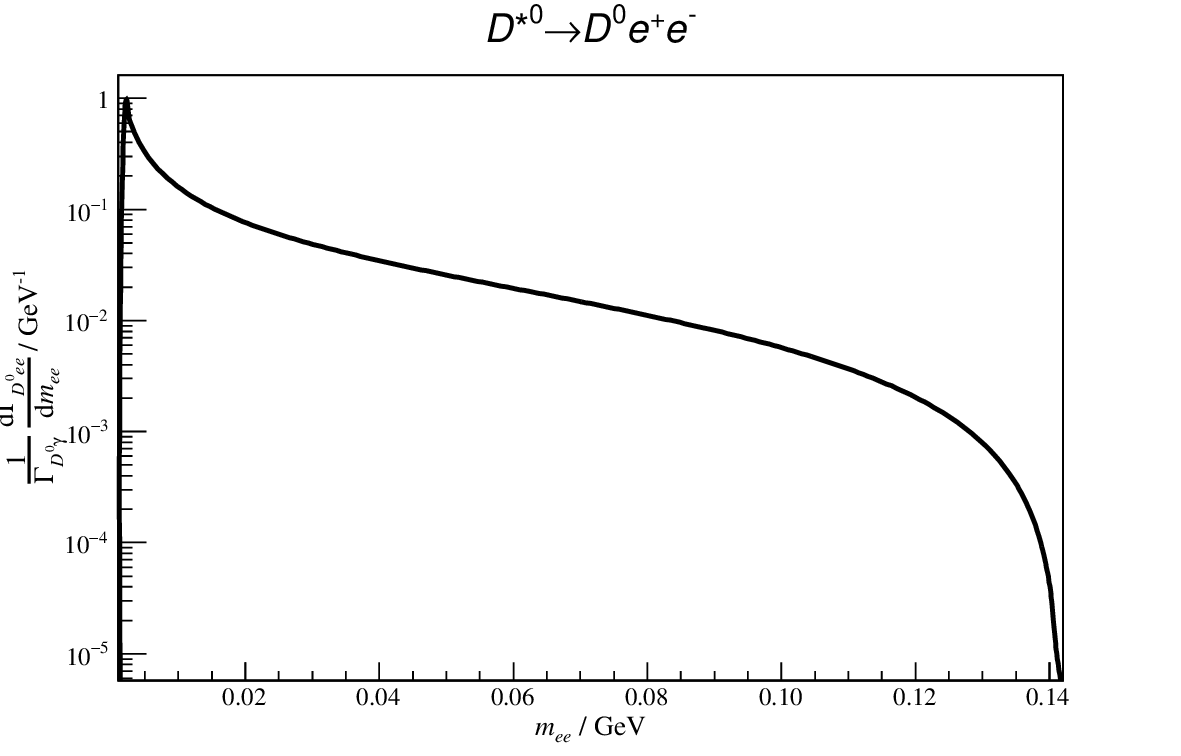}}
\end{minipage}%
\begin{minipage}{0.33\textwidth}
\centerline{\includegraphics[width=\textwidth]{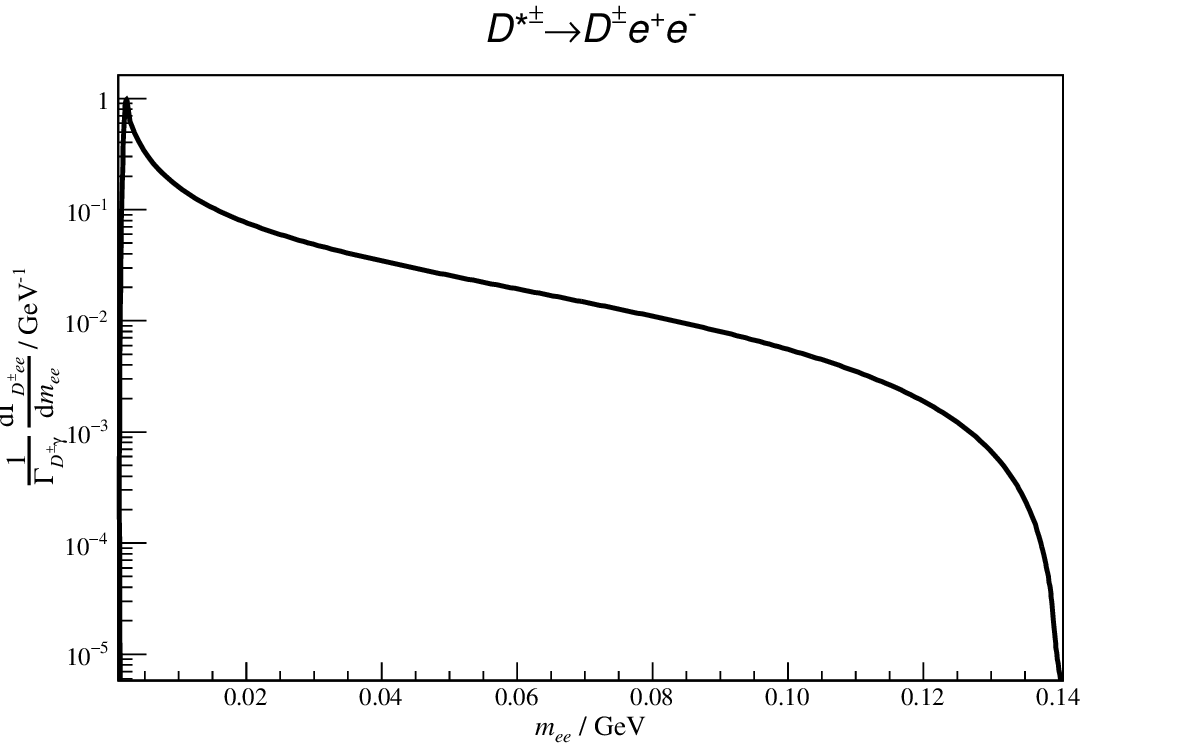}}
\end{minipage}%
\begin{minipage}{0.33\textwidth}
\centerline{\includegraphics[width=\textwidth]{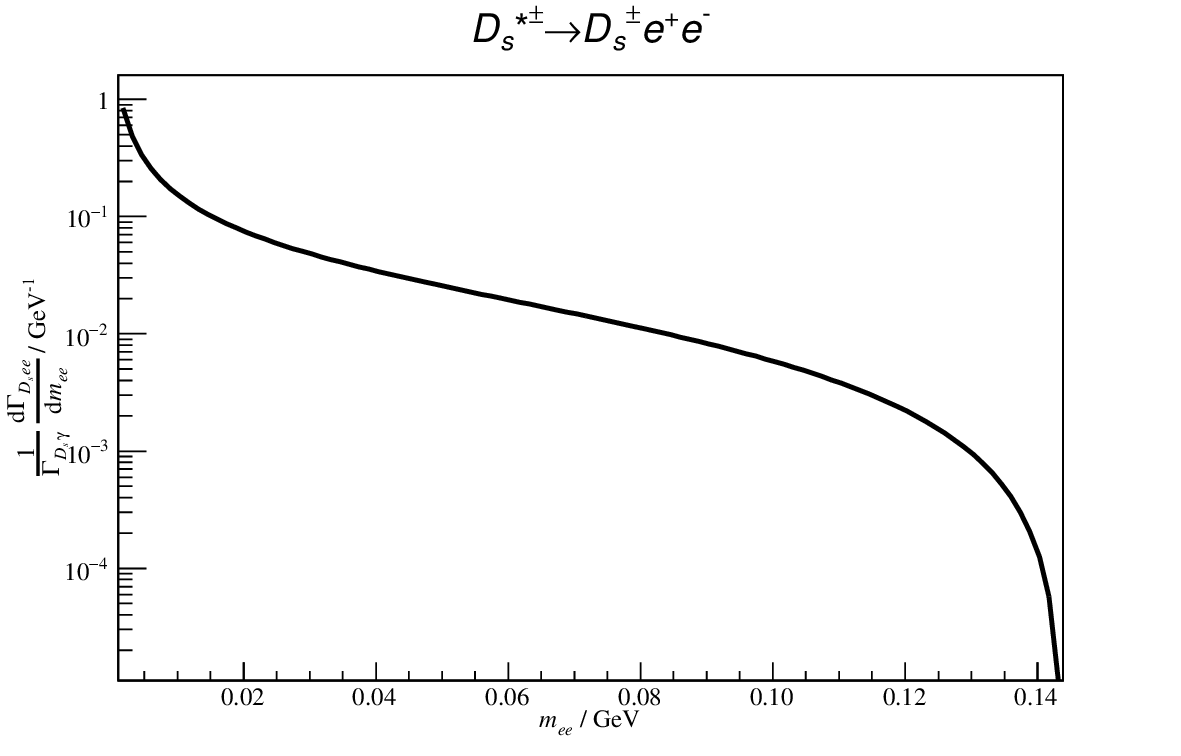}}
\end{minipage}\\
\begin{minipage}{0.33\textwidth}
\centerline{\includegraphics[width=\textwidth]{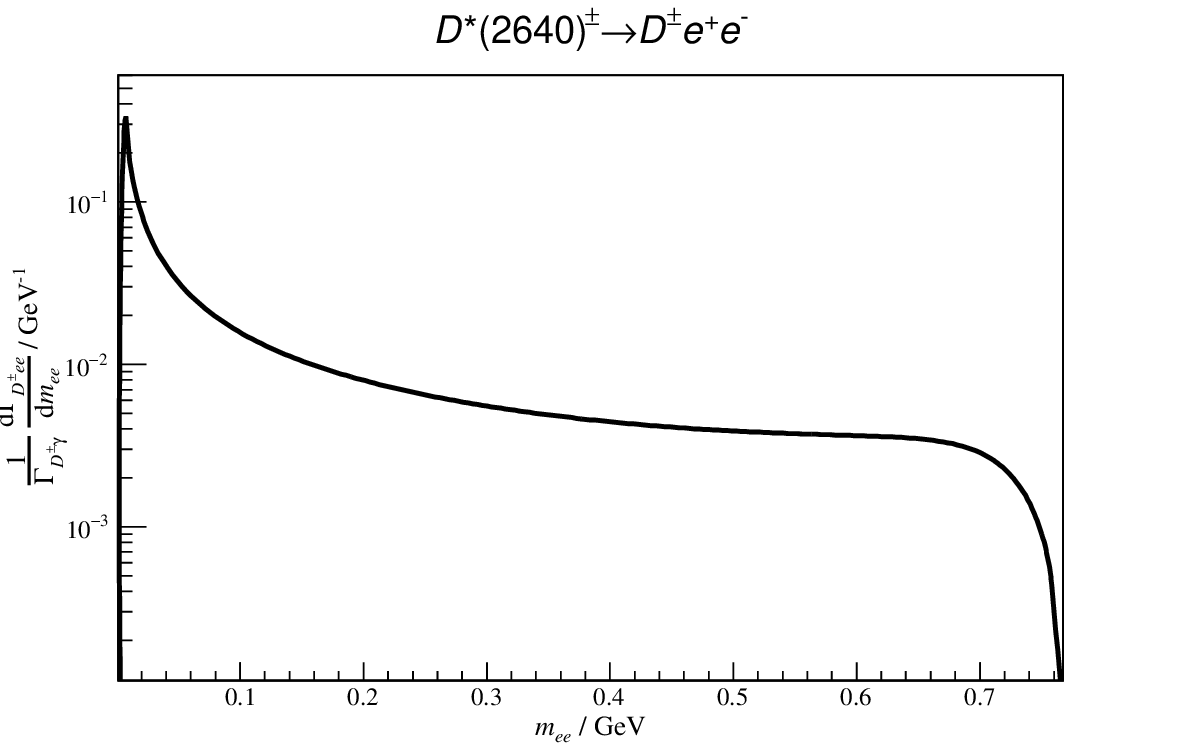}}
\end{minipage}%
\begin{minipage}{0.33\textwidth}
\centerline{\includegraphics[width=\textwidth]{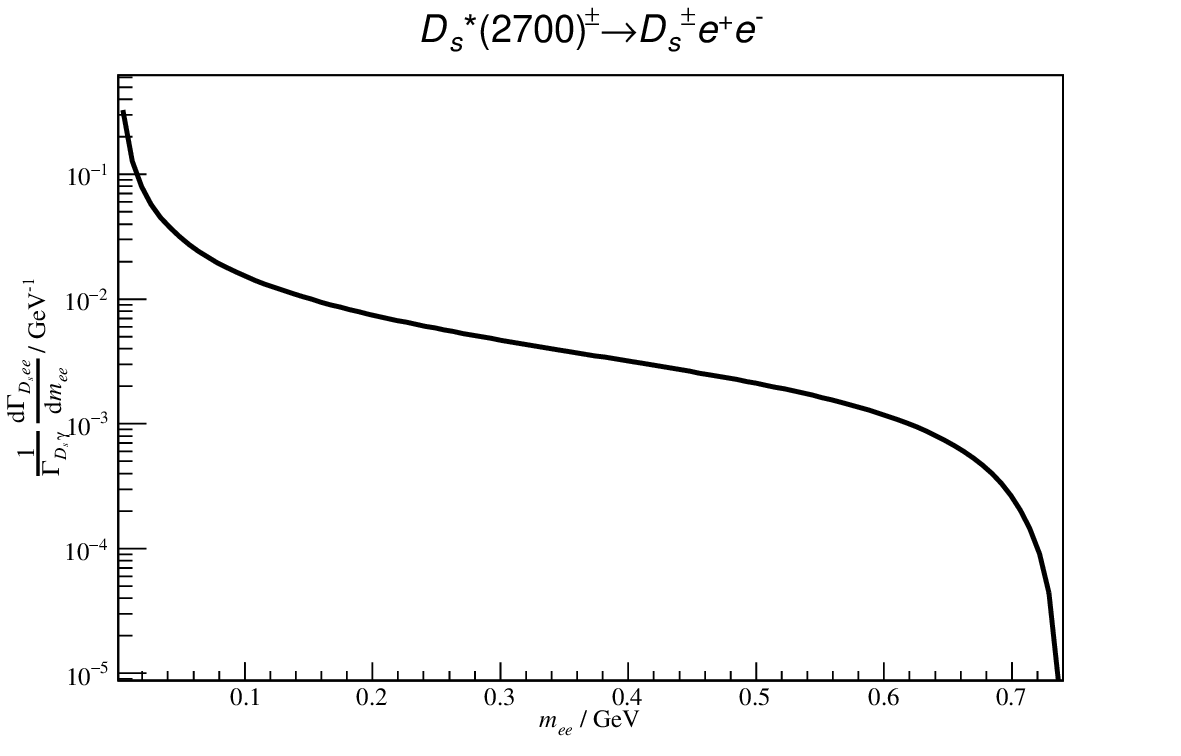}}
\end{minipage}%
\begin{minipage}{0.33\textwidth}
\centerline{\includegraphics[width=\textwidth]{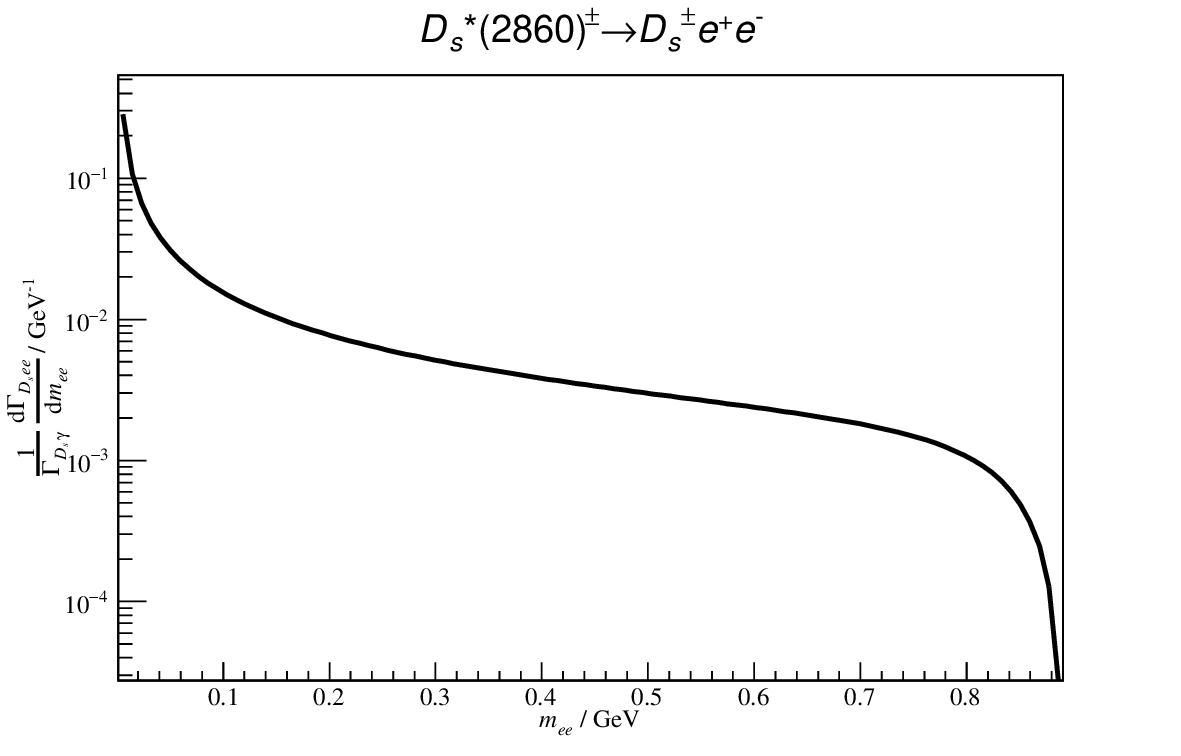}}
\end{minipage}\\
\begin{minipage}{0.33\textwidth}
\centerline{\includegraphics[width=\textwidth]{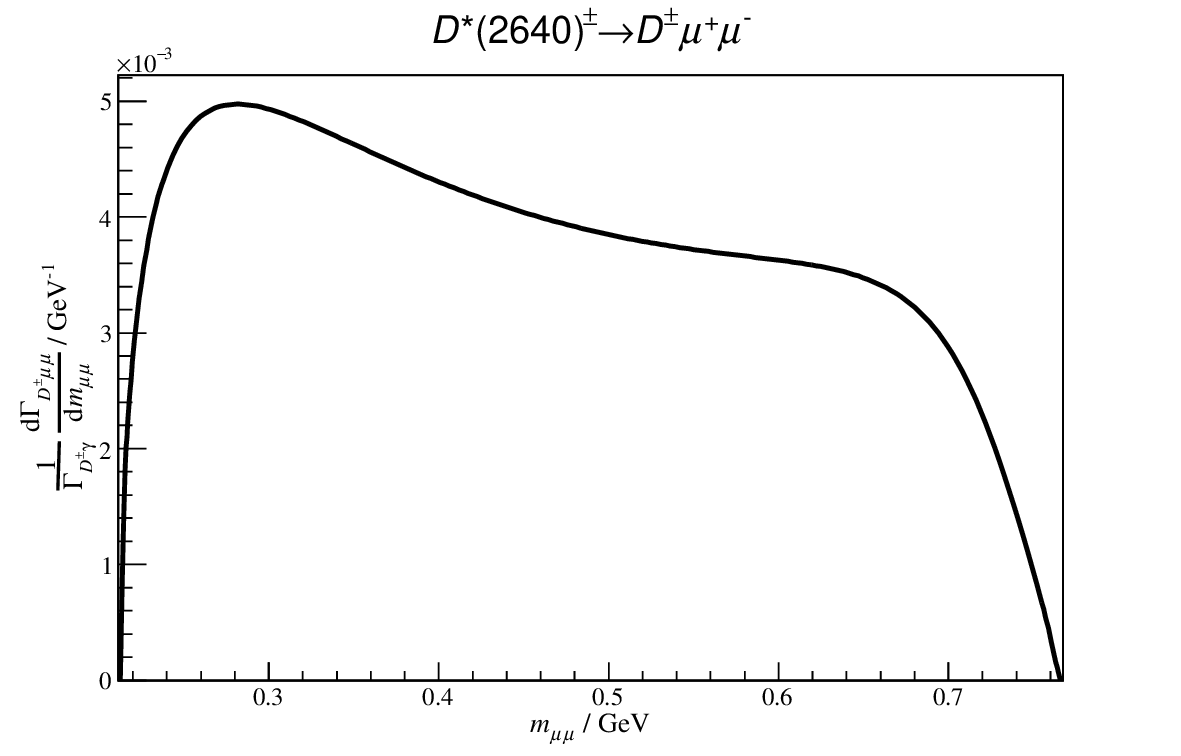}}
\end{minipage}%
\begin{minipage}{0.33\textwidth}
\centerline{\includegraphics[width=\textwidth]{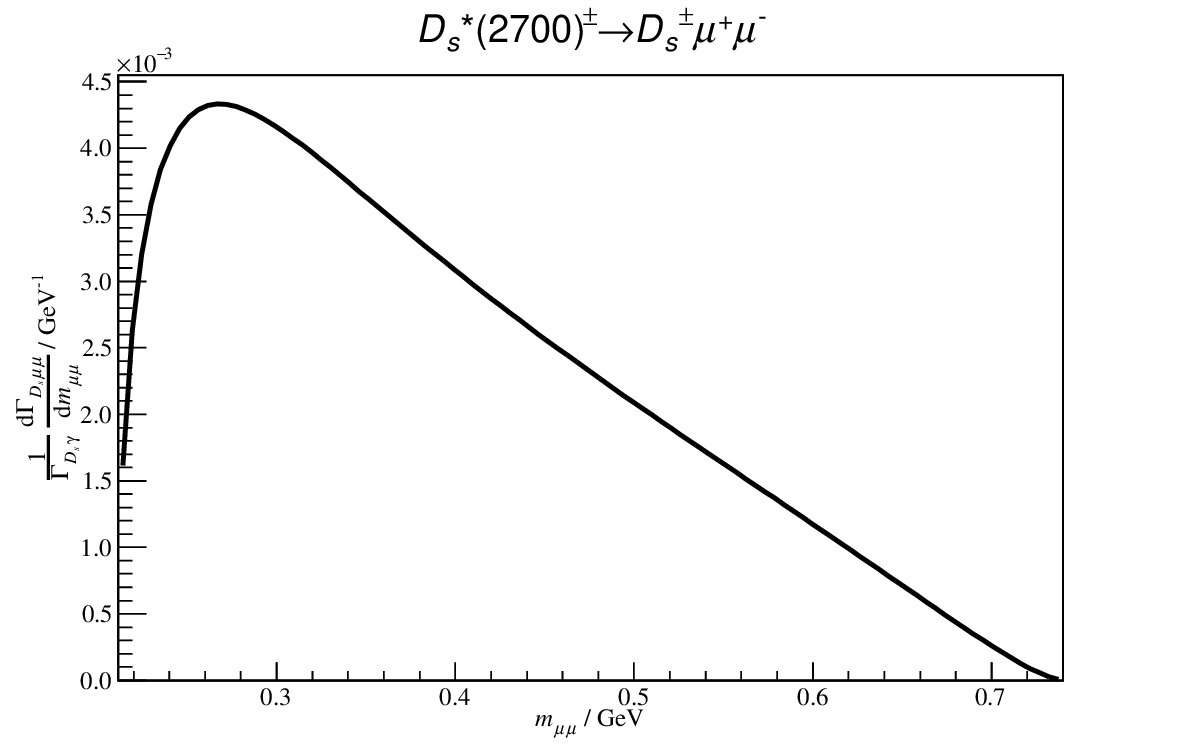}}
\end{minipage}%
\begin{minipage}{0.33\textwidth}
\centerline{\includegraphics[width=\textwidth]{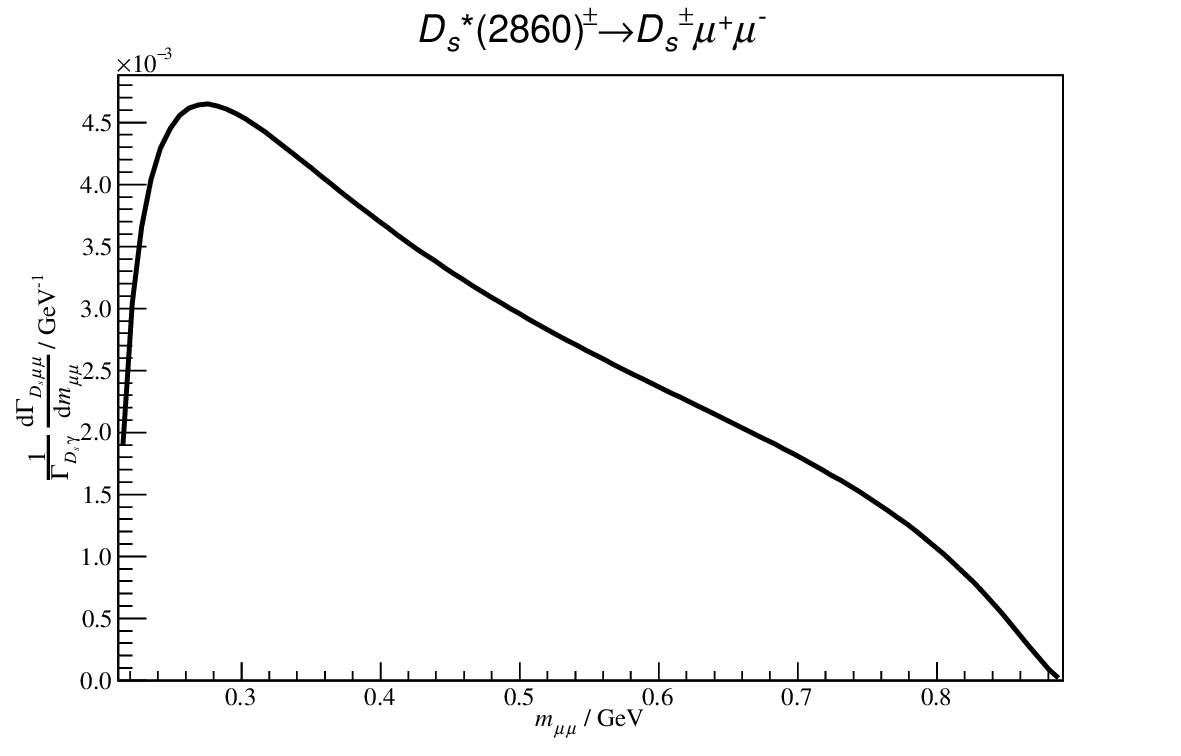}}
\end{minipage}%
\caption{The invariant mass spectra of dilepton for decays $D_{(s)}^\ast\to D_{(s)}\LL$.}\label{massspectrum}
\efg

The normalized leptonic angular distributions are also shown for $D^\ast(2640)^\pm\to D^\pm\LL$ in Fig.\ref{angulardistribution}, where $R^{\theta^\ast}_{D\ell\ell}=(\mathrm{d}\Gamma_{D^\pm\LL}/\mathrm{d}\theta^\ast)/\Gamma_{D^\pm\LL}$. The other angular distributions are almost the same when they decay with the same dilepton final states. The angular distributions between $\EE$ final state and $\MM$ final state behave differently in the range around $1-2$ radian.

\bfg[!htb]
\begin{minipage}{0.5\textwidth}
\centerline{\includegraphics[width=0.9\textwidth]{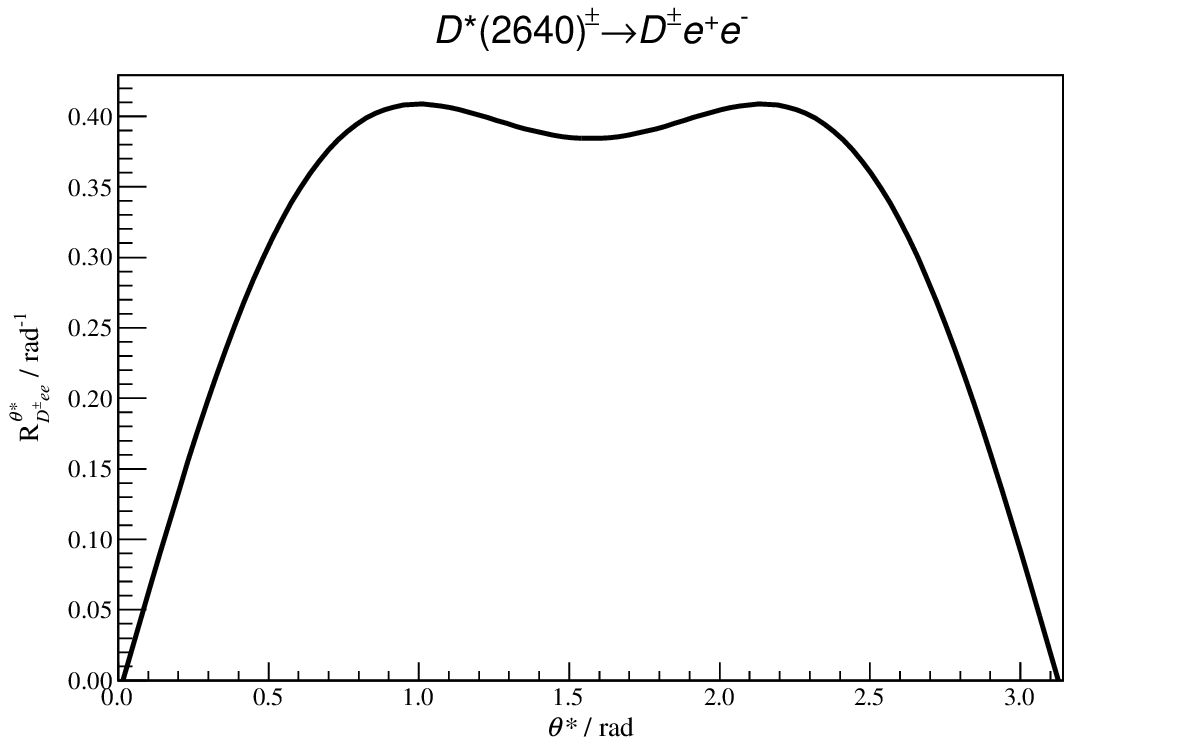}}
\end{minipage}%
\begin{minipage}{0.5\textwidth}
\centerline{\includegraphics[width=0.9\textwidth]{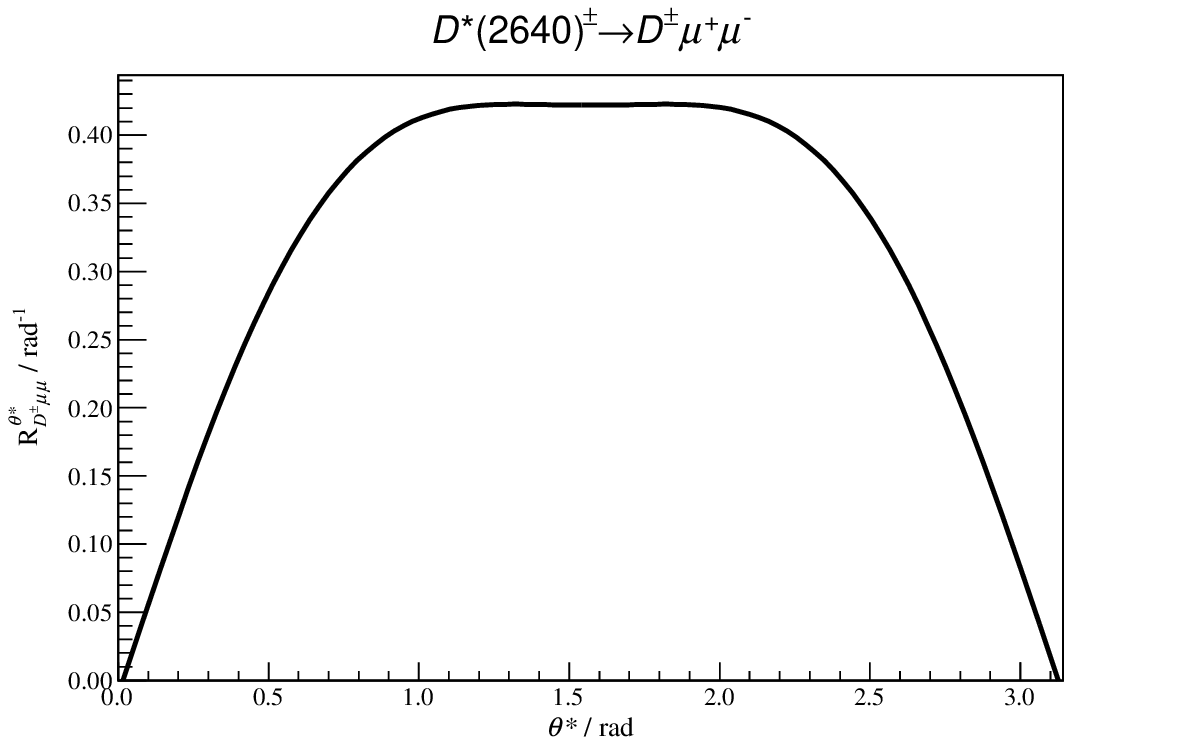}}
\end{minipage}%
\caption{The angular distributions of lepton in the rest frame of $\gam^\ast$ for $D^\ast(2640)^\pm\to D^\pm\LL$.}\label{angulardistribution}
\efg

\section{Summary and Discussions}\label{end}

In summary, we calculate the relative branching ratios of $D_{(s)}^\ast\to D_{(s)}\LL$ to $D_{(s)}^\ast\to D_{(s)}\gam$ based on TFF in VMD model, where $D_{(s)}^\ast$ represents $D^\ast(2007)^0$, $D^\ast(2010)^\pm$, $D^\ast(2640)^\pm$, $D_s^{\ast\pm}$, $D_{s1}^\ast(2700)^\pm$ and $D_{s1}^\ast(2860)^\pm$, $D_{(s)}$ represents $D^0$, $D^\pm$ and $D_s^\pm$,  $\ell$ represents electron and muon. 
 The relative branching ratios of $D^\ast(2007)^0$ and $D^\ast(2010)^\pm$ obey $\Gamma(D^\ast(2007)^0\to D^0\EE)/\Gamma(D^\ast(2010)^\pm\to D^\pm\EE)=\Gamma(D^\ast(2007)^0\to D^0\gam)/\Gamma(D^\ast(2010)^\pm\to D^\pm\gam)$, which are required by isospin symmetry. The relative branching ratio of $D_s^{\ast\pm}\to D_s\EE$ is consistent with the prediction in Ref. \refcite{cronin-hennessy2012}, as well as the experimental result\cite{cronin-hennessy2012, zyla2020}. The prediction for $D^\ast(2007)^0\to D^0\EE$ is also consistent with that in Ref. \refcite{ilten2015}. We also present the curves of magnitudes of TFFs, the invariant mass spectra of dilepton and the angular distributions of the lepton. The form factors $F^2_{D^\ast}(q^2)$ have a pole at the $m_{\rho^0/\omega}$ region. For the EM Dalitz decays with large $\Delta m$, such as $D^\ast(2640)^\pm\to D^\pm\LL$, there are enhancements around $770\mev$ in the dileptonic invariant mass spectra.
 With the BESIII $\EE\to D^\ast\bar{D}^\ast$ data samples, it is expected to measure the branching ratios of $D_{(s)}^\ast\to D_{(s)}\EE$ in experiments. The planned Super Tau-Charm Facility(STCF) can also make a measurement of the EM Dalitz decays of $D_{(s)}^\ast$. 

It is also suggested to search dark photon in $D^{\ast0}\to D^0\EE$\cite{ilten2015}. There are two vertices coupling to a dark photon and each vertex is suppressed by a factor of $\epsilon\sim10^{-4}-10^{-8}$\cite{tsai2021}, if the ordinary photon is replaced by a dark photon in the EM Dalitz decays in tree level. The decay rate of the corresponding process mediated by a dark photon will be suppressed at least by $\epsilon^4\sim10^{-16}$. Therefore, the dark photon will give little contribution to the relative branching ratios and $D_{(s)}^\ast\to D_{(s)}\gam^\ast\to D_{(s)}\LL$ can be a kind of main background in searching for dark photon.

\section*{Acknowledgments}

We thank Yin Cheng for helpful discussions. This work is supported by Joint Fund of Research utilizing Large-Scale Scientific Facility of the NSFC and CAS under Contract No. U2032114.

\appendix

\section{}

$R_\phi=0.8$  can be extracted from the ratios of decay widths of $\kaon^\ast\to\kaon\gam$. The decay width of EM radiative decay of a vector meson can be expressed as\cite{odonnell1981}:
\beq
\Gamma(V\to P\gam)=\frac{2}{3}\alpha(\frac{m_V^2-m_P^2}{2m_V})^3\frac{m_V^2+m_P^2}{2m_V^2}(\sum\mu_q)^2
\eeq
where $\mu_q=e_q/(2m_q)$ is the magnetic momenta of the component quarks. By the decays $\kaon^{\ast\pm}\to\kaon^\pm\gam$ and $\kaon^{\ast0}\to\kaon^0\gam$, we obtain that\cite{carlsmith1986}:
\beq\label{Rphi}
\frac{\Gamma(\kaon^{\ast\pm}\to\kaon^\pm\gam)}{\Gamma(\kaon^{\ast0}\to\kaon^0\gam)}=\frac{(\mu_s+\mu_u)^2}{(\mu_s+\mu_d)^2}=\frac{(R_\phi-2)^2}{(R_\phi+1)^2}
\eeq
$R_\phi=0.8$ is extracted from Eq. \eqref{Rphi} and experimental results\cite{zyla2020}.

\end{document}